# Generative Adversarial Networks for Crystal Structure Prediction

Sungwon Kim[1,†], Juhwan Noh[1,†], Geun Ho Gu[1], Alan Aspuru-Guzik[2,3,4], Yousung Jung[1,*]

[1]Department of Chemical and Biomolecular Engineering, KAIST, 291 Daehak-ro, Daejeon 34141, South Korea

[2]Chemical Physics Theory Group, Department of Chemistry and Department of Computer Science, University of Toronto, Toronto, ON M55S 3H6, Canada

[3]Vector Institute for Artificial Intelligence, Toronto, ON M5S 1M1, Canada

[4]Canadian Institute for Advanced Research (CIFAR) Lebovic Fellow, Toronto, ON M5S 1M1, Canada

† These authors contribute equally to this work.
*Correspondence: ysjn@kaist.ac.kr

# Abstract

The constant demand for novel functional materials calls for efficient strategies to accelerate the materials discovery, and crystal structure prediction is one of the most fundamental tasks along that direction. In addressing this challenge, generative models can offer new opportunities since they allow for the continuous navigation of chemical space via latent spaces. In this work, we employ a crystal representation that is inversion-free based on unit cell and fractional atomic coordinates, and build generative adversarial network for crystal structures. The proposed model is applied to generate the Mg-Mn-O ternary materials with the theoretical evaluation of their photoanode properties for high-throughput virtual screening (HTVS). The proposed generative HTVS framework predicts 23 new crystal structures with reasonable calculated stability and bandgap. These findings suggest that the generative model can be an effective way to explore hidden portions of the chemical space, an area that is usually unreachable when conventional substitution-based discovery is employed.

# Introduction

Addressing the worldwide increasing energy demand requires the discovery of novel functional materials by exploring the vast chemical space. An important subspace of chemical space is the space of crystalline materials. The essence of the successful discovery of crystal materials with desired properties depends on the exploration efficiency of chemical space. Two general strategies for this goal are either the use chemical intuition and empirical rules to improve the performance of existing materials or to search general-purpose databases of known materials, such as the experimental inorganic crystal structural database (ICSD)[1]. The latter method, known as high throughput virtual screening (HTVS)[2,3] has been demonstrated to be quite successful for various applications. Some of them include the discovery of promising photo-catalyst materials,[4,5] electrode materials for Li-ion battery,[6-8] 2D materials,[9-11] and porous materials for propylene/propane separation.[12] In these examples cited, promising materials have been identified and experimentally verified using computational screening of experimental database.

Since the currently available experimental crystal databases such as the ICSD[1] (~200,000 structural data) and the Landolt-Bornstein database[13] (6836 structural and diverse properties data) are orders of magnitude smaller than the possible chemical space of inorganic crystals, as a way to further expand the search space, the elemental substitution strategy to these known crystals is employed in many HTVS studies. Here, one performs a combinatorial elemental substitution on the existing crystal structural motifs followed by DFT calculations to generate new large computational crystal databases. Some examples of these large-scale computational databases such as Materials Project,[14] Open Quantum Materials Database (OQMD),[15] and AFLOW-lib.[16] These large computational databases have been successful in generating many new discoveries in areas such as light-harvesting materials,[17] cathode coatings of Li-ion battery using OQMD,[18] and novel antiferromagnetic Heusler compounds using AFLOW-lib.[19] Despite these promising results, one fundamental limitation of the substitution-based HTVS approach is that it cannot go beyond the template of existing crystal structures in the database.

Some of the promising methods to explore beyond the known crystal structure motifs include crystal structure prediction (CSP) methods using global optimization,[20] and generative models in machine-learning. Among various global optimization methods (e.g. basin hopping,[21] simulated annealing,[22-24] meta-dynamics,[25] minima hopping,[26] quasi-random structure search,[27,28] and evolutionary algorithm[29,30]), evolutionary algorithms are widely used in predicting crystal structures since these algorithms are population-based and can find various global and local optima with various initial guesses, and often show more robust searching without being trapped in local minima. Different evolutionary strategies[29,30] exist but generally involve two key steps, first, the initialization of structural pool (i.e. population) for the given specific chemical composition, and second, update of the population after evaluating target property (e.g. formation enthalpy) of each crystal structure using DFT calculations. Several promising results using evolutionary algorithms include the crystal structure predictions for thermodynamically stable tungsten borides,[31] Lenard-Jones cluster,[32] super-hard materials,[33] superconductors,[34] and various 2D layered materials.[35] The quasi-random structure sampling method such as *ab initio* Random Structure Searching (AIRSS)[27,28] is also noteworthy due to its simplicity in quasi-random structure generation with certain rules (e.g. symmetry, volume, and coordination) and their effectiveness to find global minimum with highly parallel implementation.

Generative models, on the other hand, focus on building a continuous materials vector space (or latent space) to encode the information embedded in the materials dataset and use the previously-constructed latent space to generate a new data point (i.e. a material). In addition, by building a mapping between the latent space and the property space, an inverse mapping of new materials with a target property can be possible. This approach is a potential solution to the long-sought goal of the community of *inverse design*.[36,37] Even without this the latent-space-property mapping, the new set of materials generated via generative models can be employed as feeder structures for a more unbiased or unstructured sampling of chemical space by means of HTVS. Since the generated materials can have completely different structures and compositions from the known materials, this generative-HTVS approach can also lead to novel discoveries that is not possible using conventional HTVS limited by the existing crystal databases. This latter approach, a crystal generative model followed by HTVS is the subject of this work.

Two of the most popular generative models in chemistry are the variational autoencoder (VAE)[38] and generative adversarial networks (GAN).[39-43] VAE typically consists of two deep neural networks (i.e. encoder and decoder), and explicitly constructs the latent space using a known prior distributions such as a Gaussian distribution. The encoder network encodes the chemical space into a low dimensional latent space, and the decoder network performs the inverse mapping that generates material structures from it. On the other hand, a GAN uses a decoder (or generator) and discriminator to learn the materials data distribution implicitly. We will further describe the framework in the **'Composition-conditioned Crystal GAN'** section. In both VAE and GAN approaches, a key component of crystal structural generative models is the invertibility from material representation (features) to real structure of material since the features generated from the latent vector should eventually inverted back to the real structure of material in order to confirm the generated material.[44]

Although many representations, such as based on fragment descriptors or graph-based encoding for crystal structures,[45,46] were proposed with great promise for predicting key properties of materials (e.g. formation energy, energy above the convex hull, band gap, bulk moduli, etc.), most of these descriptors and representations are not invertible (or have not been demonstrated to be invertible) to real 3D structure. Thus, constructing an invertible representation is still an important task for developing crystal structure generative model. One of the first suggested representations to encode crystal structures was a 3D-image representation[37] which led to the first generative model (iMatGen) for inorganic solids, which employed a VAE architecture. A similar approach was also proposed by Hoffmann et al.[47] by using 3D atomic density representations and VAE, in which an additional U-net network was employed to classify element information from the generated 3D atomic density. Kim et al.[48] proposed a WGAN-based generative model to discover new zeolite materials with desired energy and heat of adsorption. While these 3D voxel image representations opened the door to the generative modeling of the inorganic crystals, there is room for improvements for practical applications. Some of the challenges to overcome using this approach include: 1) inverting representations to materials structures requires user-defined post-processing; 2) the unit cell size of the crystal material is limited by the cubically-scaling three-dimensional grids; 3) representations are memory-intensive, leading to long training time; and finally 4) images are inherently not translational-, rotational-, and supercell-invariant.

In this work, we use a crystal representation that is inversion-free with a low memory requirement (by a factor of 400 compared to the 3D voxel representation used in iMatGen,[37] for example). We represent the crystal structure as a set of atomic coordinates and cell parameters, inspired by 'point cloud'[49-53] used for image classification and segmentation in machine learning fields, where objects are considered as a set of points and vectors with 3D-coordinates. As an application, we construct a GAN to generate new crystal structures with a desired chemical composition, and apply it to the Mg-Mn-O ternary system. The Pourbaix stability and bandgaps of these materials are then evaluated to find a promising photoanode material for water splitting in the HTVS manner.[4] The employed generative-HTVS predicts 23 novel Mg-Mn-O structures as a potential photoanode which could not have been found using the conventional substitution-based database enumeration approach.

# Representation

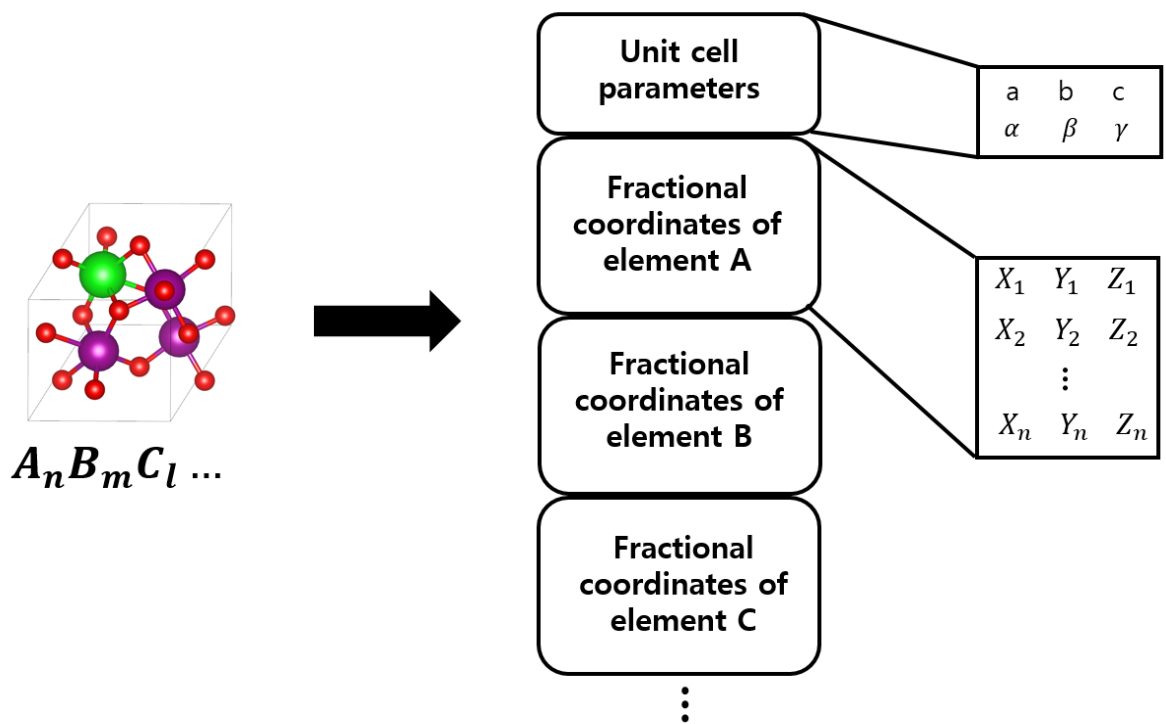

**Figure 1. Point cloud representation of crystal structure.** The representation is composed of unit cell parameters and the sets of rescaled fractional coordinates of atoms.

To encode the crystal structure, we employ a 2D matrix representation inspired by a 'point cloud'[50] which includes both unit cell and fractional coordinates of each atom in the unit cell where the permutational invariance is imposed by symmetry operation used in network encoding the proposed 2D representation (see **'Composition-conditioned Crystal GAN'** section for model detail). Since the representation is the material structure itself, there is no need for the inversion from the representation to the material. One limitation is the lack of translational, rotational, and supercell invariances (i.e. invariance under the repeating of the unit cells with respect to the lattice vectors) of the representation, and we address them by data augmentation as outlined later. The representation is summarized graphically in **Figure 1**. Since our representation only requires the atomic coordinates and cell information, it requires almost no preparation and memory cost to store the raw input data, in contrast to the 3D voxel representations which require substantial memory space to store the grid data.

We note that a similar representation was recently used to generate new ternary hydride structures by learning their binary counterparts with a cross-domain learning strategy.[54] Interestingly, the method generated the structures of more complex domain with reasonable interatomic distances by imposing constraints in the training process. However, it differs from our work in that it is a cross domain model; generating structures of more complex domain (ternary) from the structures of less complex domain (binary).[55]

**Training dataset and data preprocessing**

As mentioned previously, as an application of the proposed GAN model for crystal structure generation, we considered the ternary Mg-Mn-O system to generate new crystal structures of various compositions. The training set for the Mg-Mn-O system was constructed using the elemental substitution of the ternary compounds in the Materials Project (MP) database.[56] After removing duplicates, we retain a total of 1240 unique structures with 112 compositions in the initial training set. We note that this dataset has the data imbalance in the composition and affine invariance issues such as supercell, translation, and rotation. To address them, we used data augmentation, which is a commonly used technique in the machine learning field to alleviate such data imbalance and invariance problem.[57-61] Specifically, we added the supercell structures as well as the structures in which translational and rotational (i.e. swapping the axes of the unit cell) operations are applied until these augmentations yield 1,000 structures for each composition. Since the original training dataset includes 112 Mg-Mn-O compositions, a total of 112,000 Mg-Mn-O structures were used for the training of the current generative model. In addition, for the robust training of the classifier, when the training data was put in the models, atomic permutation operations were randomly applied to training data. Information for V-O dataset is described in section **S6** in **SI**. The learning curve of the Composition-conditioned Crystal

GAN and the effects of data augmentation for addressing symmetry invariance are described in section **S3** in **SI** and section **S7** in **SI,** respectively. Compared to a model without data augmentation, the analyses in **Figure S11** show that data augmentation clearly improves the model's ability to recognize the same materials represented in different input features (translated, rotated, or supercell repeated) as identical.

## Composition-conditioned Crystal GAN

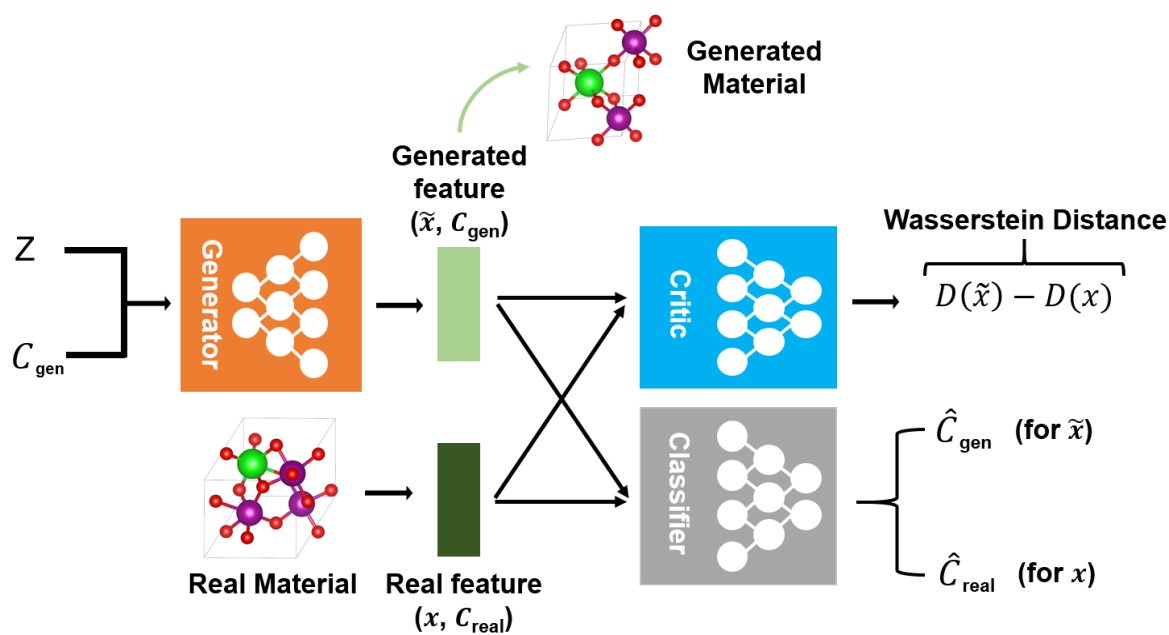

**Figure 2. Composition-conditioned Crystal GAN proposed in this work for inorganic crystal design.** Z, $C_{gen}$, and $C_{real}$ denote a random input noise, user-desired composition condition, and composition of real material, respectively. The variables $\tilde{x}$ and $x$ denote the feature (representation) of generated and real materials, respectively. $\hat{C}_{gen}$ and $\hat{C}_{real}$ denote the predicted composition of the generated and real features, respectively. D($x$) is the critic function also known as critic network.

Our GAN model consists of three network components: a generator, a critic and a classifier as shown in **Figure 2**. The generator takes random Gaussian noise vector ($Z$) and one-hot encoded composition vector ($\boldsymbol{C_{gen}}$) as input to generate new 2D-representations. The one-hot encoded composition vector is used as a condition to generate materials with target composition. The critic computes Wasserstein distance which represents dissimilarity between the true and trained data distributions, and by reducing this distance the generator would generate more realistic materials. The critic network is composed of three-shared multi-layers perceptions (MLP) followed by average pooling layers to ensure the permutation invariance under the reordering of points in 2D-representation.[50] We note that the permutation invariance under the reordering of input is satisfied by using shared weight parameters and average pooling since the averaged value is unchanged under the change of orders. The classifier network, which outputs composition vector from the input 2D-representation, is used to ensure that the generated new materials meet the given composition condition. The loss of the classifier is back-propagated to the generator only if the generated 2D-representation ($\tilde{x}$) is taken as input. More details on the architecture of each neural network, hyper-parameters for the model, and loss function are described in section **S2** of **Supporting Information**, **SI**.

# RESULT AND DISCUSSION
## Comparison with iMatGen

Before applying the current model to the Mg-Mn-O system, we first compared the results on the V-O system that was employed in the iMatGen[37] work, which represents the first generative model for inorganic crystal structures and therefore it is an useful baseline to explore. After using a data-augmented version of the V-O training data, we generated samples of $V_3O_4$, $V_4O_5$, $V_5O_6$, $V_5O_8$, and $V_6O_7$ structures to compare the chemical space generated from the iMatGen based on VAE. About 40% of the metastable polymorphs of V-O ($E_{hull} \leq 200$ meV/atom) discovered by iMatGen were re-discovered by the current GAN model, indicating some similarity in the latent space trained by each generative model. The remaining 60% difference in the two (VAE and GAN) generative models can thus be interpreted as a difference in the latent space structure or sampling method in each generative model. Particularly, in the $V_3O_4$ and $V_6O_7$ composition, the present model generated more stable polymorphs than the most stable ones generated via iMatGen. Thus, the performance of the current coordinate-based GAN model seems comparable to that of iMatGen. Given that the current model can sample the compounds with user-desired composition with various invariances also addressed for larger crystal unit cell, it can be particularly useful for discovering materials with specific compositions. The other training details and the results for the V-O system are summarized in the section **S6** in **SI**.

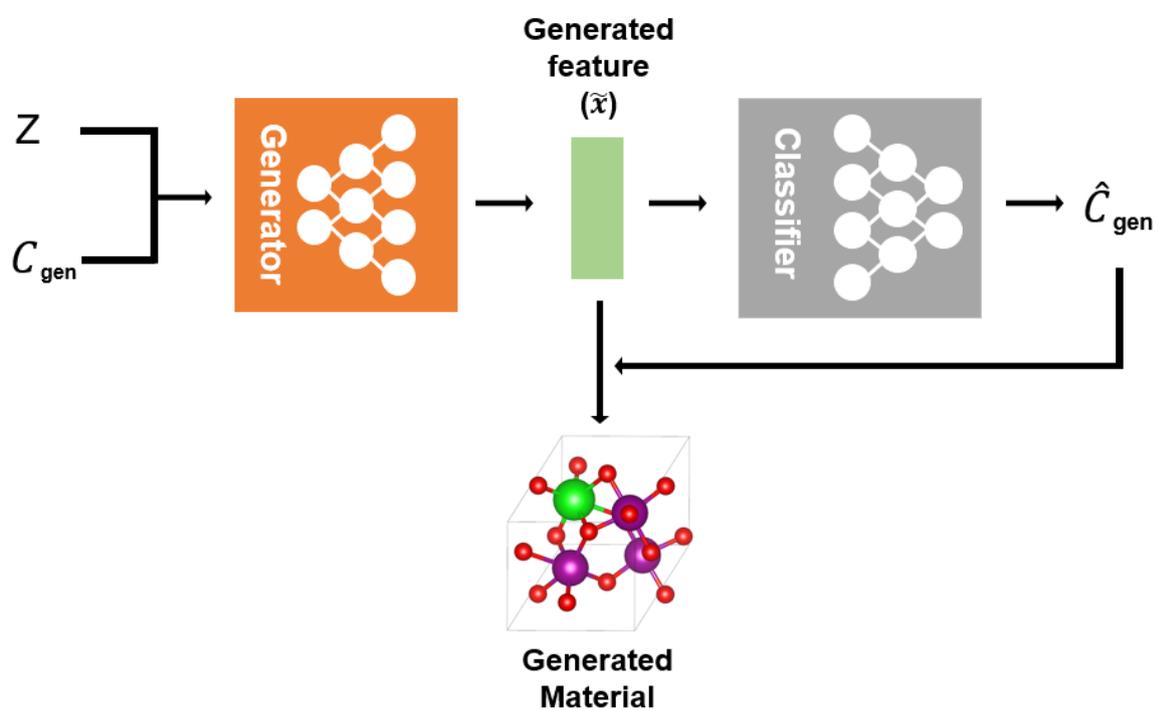

**Figure 3. Schematic of the generation process for crystals with the desired composition**. The composition of generated material is determined by the output of the classifier network.

# Generative high-throughput screening of ternary Mg-Mn-O photoanode materials

We generated ternary Mg-Mn-O materials and evaluated their photoanode properties to find structures with an improved performance. A previous study[4] demonstrated that Mn oxides combined with Mg resulted in reasonable catalytic activity but with relatively weak aqueous stability in experimental conditions (pH and voltage). Thus, to further enhance the aqueous stability, a computational HTVS study based on an elemental substitution of the MP database (total 7,356 candidates) was previously performed which resulted in a new discovery of $Mg_2MnO_4$ with reasonable stability and activity (also experimentally verified).[56] In this work, we apply the proposed generative model to perform generative-HTVS to find new Mg-Mn-O structures beyond the existing structural motifs in the database. To achieve this, first we set total 133 candidate compositions (see **Figure 4b**) that meet the condition of Mn oxidation state ($2 \leq OS_{Mn} \leq 4$), which are expanded from the chemical space consisting of existing materials (see **Figure 4a**). Among 133 compositions, we selected total 31 compositions (11 compositions included in MP, and 20 compositions not included in MP) by considering the number of atoms in the unit cell due to the computational cost of DFT. Then, we sampled total 9,300 Mg-Mn-O structures using the proposed crystal GAN; 3,300 structures (300 structures in 11 compositions included in MP, see **Figure 4c**), and 6,000 structures (300 structures in 20 new compositions not in MP, see **Figure 4d**). The process of sampling materials is described in **Figure 3**. These generated crystal structures are then fed to the DFT calculations for property evaluation.

The energy above hull (formation stability) of the generated materials is first summarized in **Figure 4c**. Among the 3,300 newly generated materials for the existing compositions in MP (**Figure 4c**), 368 Mg-Mn-O materials are predicted as theoretically meta-stable (i.e. $E_{hull} \leq 200$ meV/atom, red crosses in **Figure 4c**) where 35 structures are considered as potentially synthesizable[62] (i.e. $E_{hull} \leq 80$ meV/atom). Among those 368 newly generated materials with $E_{hull} \leq 200$ meV/atom, 60 of them are the same as those discovered by the previous HTVS on the 7,500 substituted dataset[56]. In particular, for the MgMn$_4$O$_8$ composition, the current model-generated structure is very close to the convex hull (i.e. $E_{\text{hull}} = 5$ meV/atom), much more stable than all the related polymorphs found in MP. This shows that the present crystal generative model can discover new stable compounds missed out by conventional substitution-based methods.

The formation stability for the compositions that are not in the MP database is next summarized in **Figure 4d**. Among the 6,000 generated structures, 753 Mg-Mn-O materials are predicted as theoretically meta-stable (i.e. $E_{\text{hull}} \leq 200$ meV/atom, red crosses in **Figure 4d**) where 113 structures are considered as potentially synthesizable (i.e. $E_{\text{hull}} \leq 80$ meV/atom). In particular, for Mg$_2$MnO$_4$, a composition not in MP, we discovered a structure corresponding to the convex hull minimum indicating that our model can discover an entirely new ground state material within the DFT accuracy.

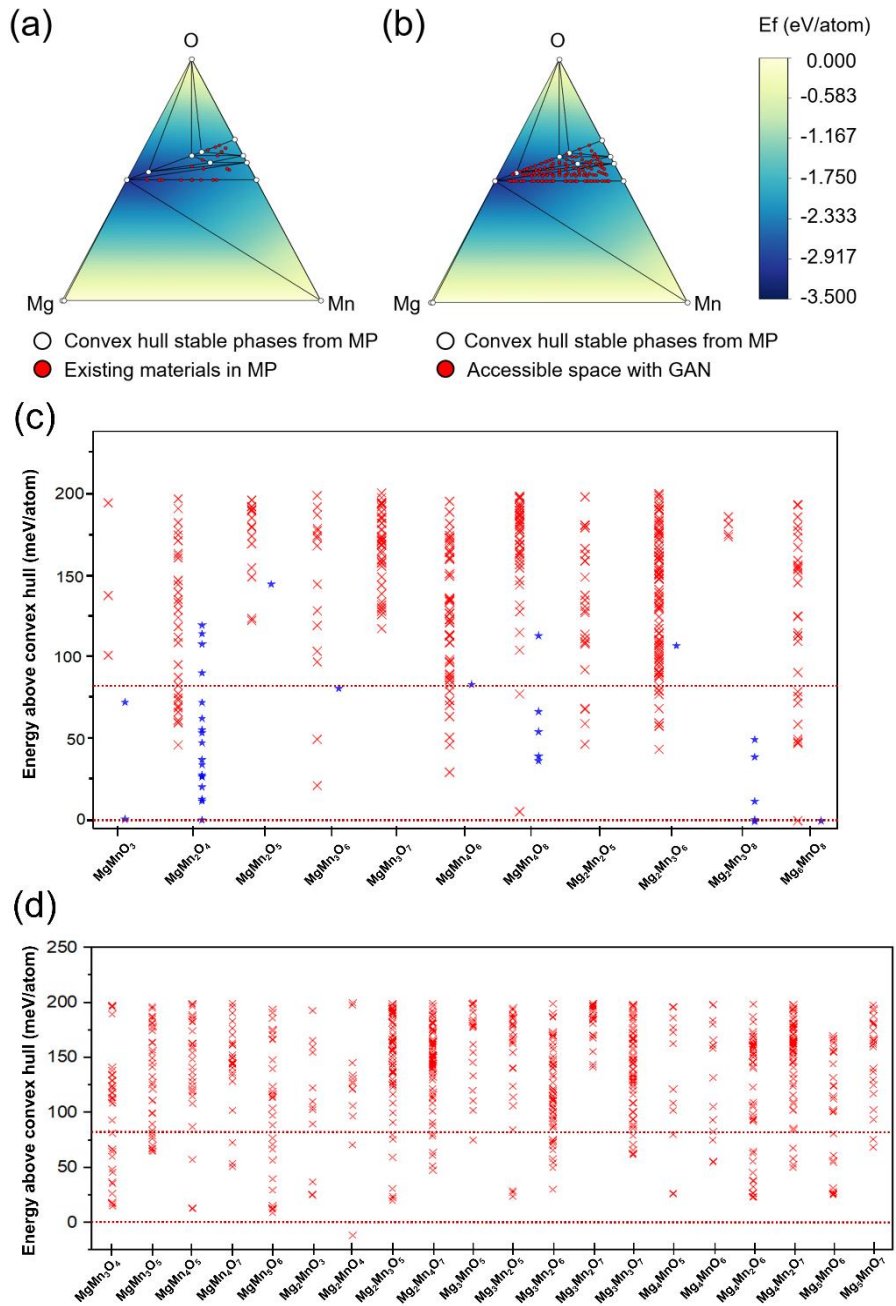

**Figure 4. Phase diagram and DFT calculated stability for the generated Mg-Mn-O materials**. Ternary phase diagram of Mg-Mn-O system constructed using the convex hull stable phases taken from the materials project database (green circle), including (a) metastable Mg-Mn-O compositions (red circle) taken from materials project or (b) possible compositions that can be explored by our proposed generative model. Stability of the crystal structure in form of the energy above the convex hulls is computed using DFT for (c) 11 compositions included in the MP database, and (d) 20 new compositions not in the MP database. Red crosses are the generated materials with composition-conditioned, and blue stars in (c) correspond to the materials in the MP database. (There are no meta-stable ($E_{hull} \leq$ 200meV/atom) structures having $Mg_2Mn_2O_5$ composition in MP database.) The horizontal dotted red lines represent 80meV/atom and 0meV/atom, respectively.

Since Mg-Mn-O compounds are considered here as photoanode materials, their Pourbaix stability ($\Delta G_{pbx}$) and the band gaps ($E_g^{HSE}$) are further considered as the next screening criteria for those newly generated structures that satisfy $E_{hull} \leq 80$ meV/atom (35 materials in **Figure 4c** and 113 materials in **Figure 4d**). The Pourbaix hull represents the stability of a material in an aqueous electrochemical environment at a given pH and electrochemical condition[63] (i.e. difference of the free energy from the ground state). We evaluated such aqueous electrochemical stability described by the minimum of Pourbaix hull Gibbs free energy at 1.5 V *vs.* RHE over the 0-14 pH range, $\Delta G_{pbx}^{min}$, which was calculated as implemented in the Pymatgen[64] module (also refer to Noh et al.[56] for computational details). Therefore, a material with low $\Delta G_{pbx}^{min}$ represents a (meta-)stable phase in an aqueous electrochemical environment, and for those materials meeting $\Delta G_{pbx}^{min}(E_{form}) \leq 0.8$ eV/atom, the HSE calculations are further performed to calculate the band gap.

Following Shinde et al.[4], we finally identified 28 Mg-Mn-O materials (**Figure 5**) with $\Delta G_{pbx}^{min}(E_{form}) \leq 0.59$ eV/atom and 1.6 eV $\leq E_g^{HSE} \leq$ 3.0 eV as potential photoanode material. Out of these 28 Mg-Mn-O materials, 14 materials are corresponding to new compositions not included in database, meaning those are entirely new structures. Remaining 14 materials are composed in 8 existing compositions in the database, among which 5 of them correspond to the previous findings by Noh et al.[56] based on substitutional HTVS; we have used the Structure Matcher function in Pymatgen python package to estimate the structural similarity, and more detailed discussion is described in **S5.2 in SI**. Experimentally, in $MgMn_2O_4$,[4] $Mg_6MnO_8$,[65] and $Mg_2MnO_4$[56] compositions, promising photoanode materials were synthesized. We found several promising photoanode materials in many other compositions which could not be considered in conventional HTVS. (See **Figure S9**) Some of the 23 newly found photoanode candidates (14 materials in new compositions, and 9 materials in existing compositions) are depicted in the section **S5.3** in **SI**.

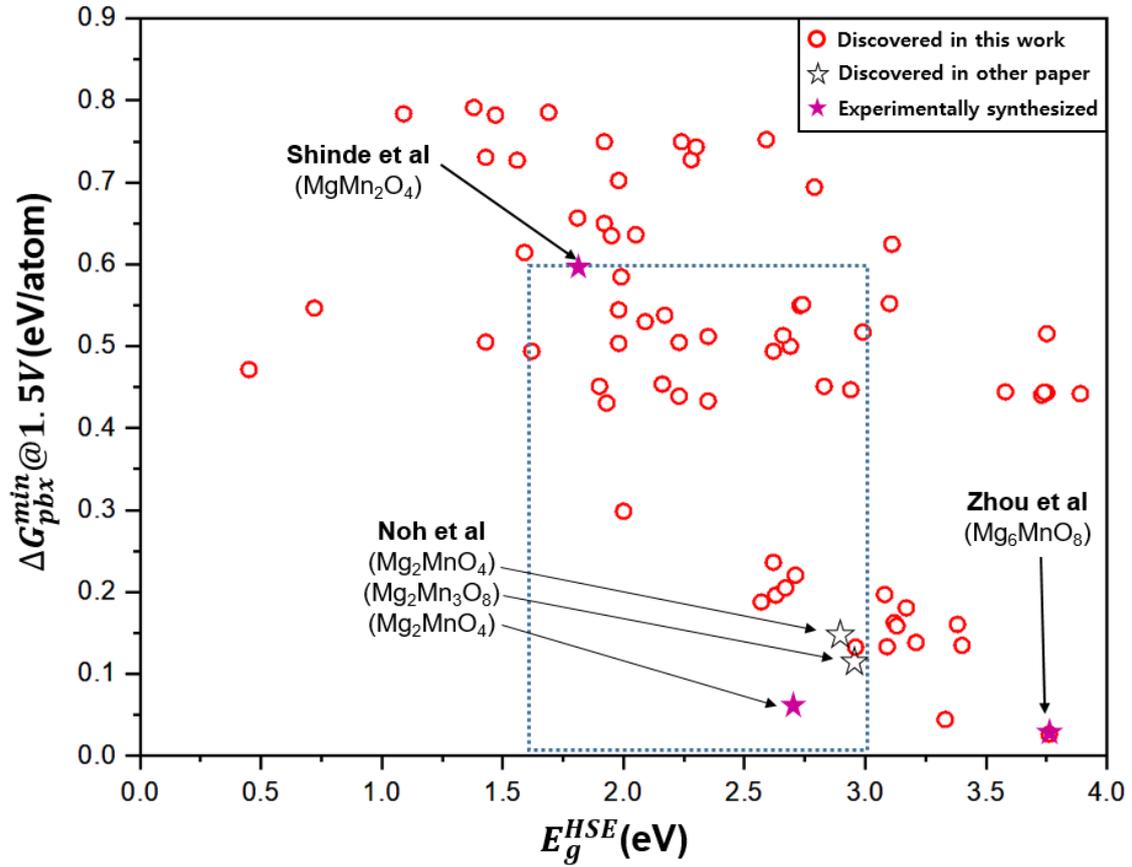

**Figure 5. Pourbaix stabilities and HSE-bandgap energies of stable structures generated by the proposed crystal GAN model (red circles).** The dashed blue box is the target region for promising photoanode material. Stars represent the promising photoanode materials discovered by other previous works (i.e conventional HTVS)[4,56,65] and purple stars are materials synthesized experimentally.

# Discussion

The proposed generative framework can be compared with crystal structure prediction methods using evolutionary algorithms[29,30] and quasi-random searching (i.e. AIRSS[27,28]). As briefly described in Introduction, evolutionary algorithms search an optimal state (material) by repeating the series of specific evolutionary processes rather than learning the distribution of the whole target chemical space as in GAN. The quality of the results (e.g. how close the final structure is to the global minimum and how diverse the local minimum structures are) and computational cost to obtain optimal state might be sensitive to this initialization in case of exploring entirely new chemical space as evolutionary algorithms start from randomly initialized population. In case of quasi-random searching approach[27,28], it randomly samples the structures to maximize the exploration but usually steered by human-intuitive constraints, such as symmetry and coordination numbers, towards more realistic structures. In general, large computational cost to find new materials would be a main challenge of most global optimization-based strategies, so there have been additional efforts to reduce computational cost of evaluating property by assisting or replacing the *ab initio* approach via the property predictive machine learning models.[66]

Compared to the aforementioned global optimization strategies which explore new local minima by utilizing the previous trajectories on the configurational space (i.e. on-the-fly approach), the generative framework generates new data (material) from the continuous latent space that encodes the information of the entire chemical space used in training stage. This means that the efficiency and accuracy of structure prediction is largely dependent on the structural diversity of the training dataset. Of course, the computational cost to prepare the training dataset and optimize the generated structures is also a burden for the present generative model-based prediction as in most other global optimization techniques. Thus, the methods based on global optimization and the generative-HTVS seem comparable and complementary in the sense that the former is efficiently searching for a global minimum by learning the geometric information of the potential energy surface (or functional manifold) with specific structure generation rules, while the latter is learning the whole distribution of crystal structures in the training dataset and then sample the new data from this machine-learned distribution.

There are several limitations and promising directions for the proposed composition-based generative framework to be used as general-purpose inverse design. The current model generates new crystal structures with only the target composition conditioned, and thus subsequent HTVS of properties are required to make a final functional discovery. To be truly inverse design in which the machine generates the functional material directly without HTVS, one thus should add to the composition other materials properties (e.g. bandgap energy, dielectric constant, and etc.) as input conditions to guide the materials discovery. Another way of achieving the inverse design goals would be to combine the generative process with reinforcement learning.[40] In addition, while the current model can produce ternary crystal compounds, extending it to quaternary and higher-order compounds would be straightforward by adding more rows or channels in the input format, or by separately adding a segmentation network to classify elemental information (although preparing the training data for higher-order compounds would be more challenging due to a combinatorial complexity when including more than 4 elements). Other important aspects in need of further developments is the quantitative metrics related to the novelty of generated samples compared to the existing data, as well as the uncertainty (or validity) of the generated data. Synthesizability prediction of the newly generated materials would also be an essential ingredient for practical inverse design of crystals for experimental verification.

# Conclusions

We proposed to employ the generative adversarial network (GAN) for crystal structure generation using a coordinate-based (and therefore inversion-free) crystal representation inspired by point clouds. By conditioning the network with the crystal composition, our model can generate materials with a desired chemical composition. As an application, we applied it to generate new Mg-Mn-O ternary compounds to find potential photoanode materials and discovered 23 new crystal compounds with reasonable stability in an aqueous environment and bandgap. Two of the structures (in $MgMn_4O_8$ and $Mg_2MnO_4$) corresponded to the convex hull minimum, a stable new phase, or very close to it within the DFT accuracy. We expect that the proposed model can be extended to a general-purpose inverse design by incorporating materials properties into the model in future work.

# Acknowledgement

We acknowledge generous financial support from NRF Korea (NRF-2017R1A2B3010176).

# Supporting Information

# Generative Adversarial Networks for Crystal Structure Prediction

Sungwon Kim[1,†], Juhwan Noh[1,†], Geun Ho Gu[1], Alan Aspuru-Guzik[2,3,4], Yousung Jung[1,*]

[1]Department of Chemical and Biomolecular Engineering, KAIST, 291 Daehak-ro, Daejeon 34141, South Korea

[2]Chemical Physics Theory Group, Department of Chemistry and Department of Computer Science, University of Toronto, Toronto, ON M55S 3H6, Canada

[3]Vector Institute for Artificial Intelligence, Toronto, ON M5S 1M1, Canada

[4]Canadian Institute for Advanced Research (CIFAR) Lebovic Fellow, Toronto, ON M5S 1M1, Canada

† These authors contribute equally to this work.

*Correspondence: ysjn@kaist.ac.kr

# Table of contents

# S1. Representation details

## S1.1 Specific representation of Mg-Mn-O system and rescaling fractional coordinates

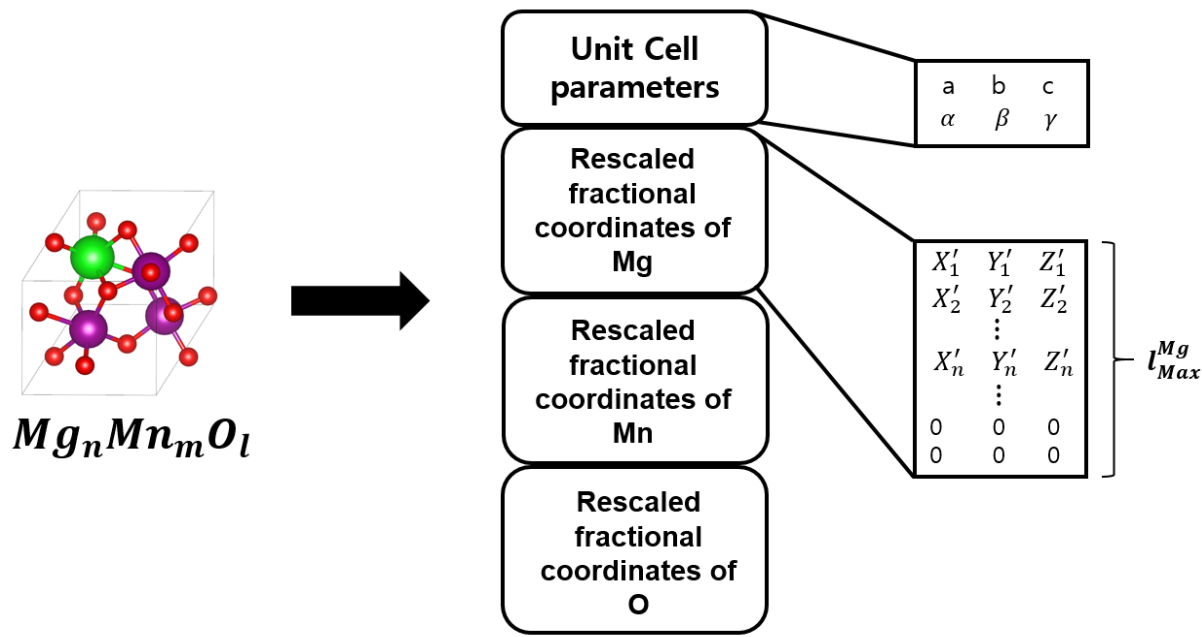

**Figure S1. Specific representation of Mg-Mn-O structure.**

The representation is composed of unit cell parameters and the sets of rescaled fractional coordinates of atoms. Each set of rescaled fractional coordinates represents an element. If the number of elements increases, the number of sets of coordinates also increases. The length of each part of the coordinates is fixed to $l_{Max}^{element}$ and the rows where atom does not exist are filled with zero-paddings

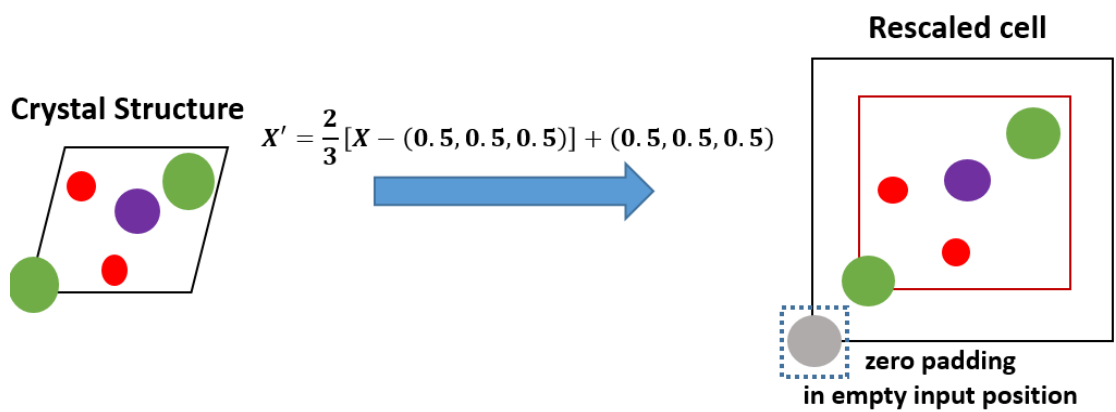

**Figure S2. Scheme of rescaling fractional coordinates.** To distinguish real atoms located at (0,0,0) from zero-paddings which were used to fix the shape of the representation, we rescaled

the fractional coordinates.

The first two rows of the 2D-representation contain the lengths of the unit cell edge and angles between them, and the fractional coordinates of atoms are listed below. To fix the shape of the representation, we set the maximum number of atoms in each element, $l_{Max}^{A}$ and zero-padding (0, 0, 0) is used if the number of atoms with specific element type A is lower than $l_{Max}^{A}$. In the Mg-Mn-O system used here, for example, $l_{Max}^{Mg}$ is 8, $l_{Max}^{Mn}$ is 8 and $l_{Max}^{O}$ is 12. (See **Figure S1**) In addition, as shown in the **Figure S2**, we apply the rescaling operation to fractional coordinates of the point atom, $\boldsymbol{P}(X,Y,Z)$ to distinguish the zero-paddings from the atoms located at (0, 0, 0) position.

$$\boldsymbol{P}' = \frac{2}{3}[\boldsymbol{P} - (0.5, 0.5, 0.5)] + (0.5, 0.5, 0.5)$$

where $\boldsymbol{P}'$ is the rescaled fractional coordinates used in our representation. Since our representation only requires the atomic coordinates and cell information, it requires almost no preparation and memory cost to store the raw input data that can be compared with the 3D voxel representations necessitating substantial memory space to store grid data.

**S1.2 Data augmentation**

- **Supercell operation**

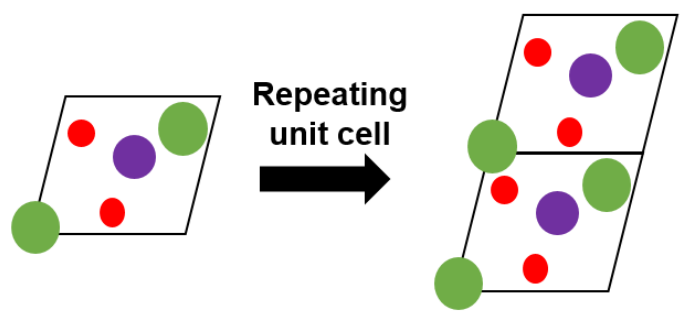

**Figure S3. Scheme of supercell operation** We made super cells by repeating unit cells twice in x-y axis, y-z axis, and x-z axis respectively where the number of atoms in the cell does not exceed the maximum number of atoms our representation.

- **Translation operation**

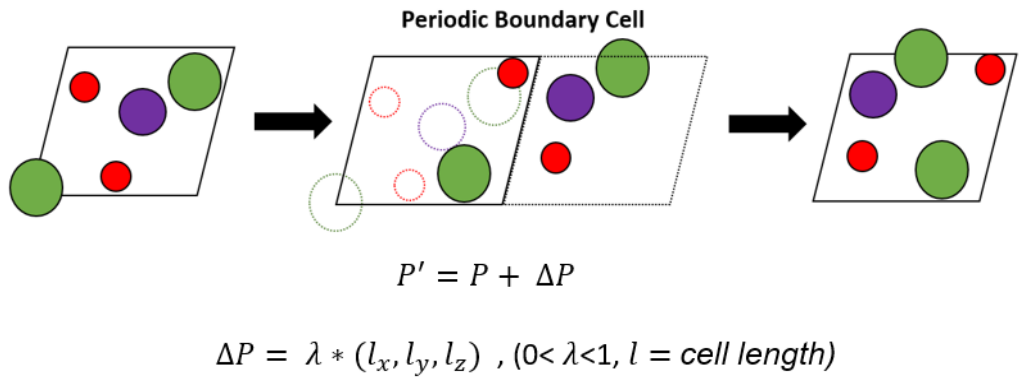

$$P' = P + \Delta P$$

$$\Delta P = \lambda * (l_x, l_y, l_z) \text{, } (0< \lambda <1, l = \text{cell length})$$

**Figure S4. Scheme of translation operation** We applied translation operation on structure data by moving atoms in unit cell by random distances which are smaller than cell length.

- **Rotating operation**

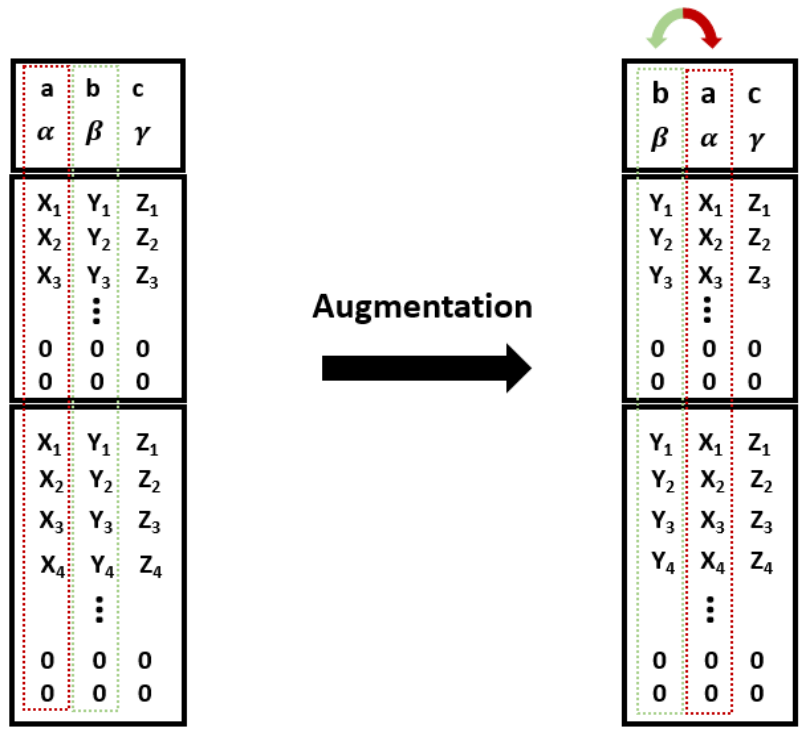

**Figure S5. Scheme of rotating operation** We applied rotational operation on structure data by swapping two axes (column).

## S2. Model details

### S2.1 Architecture of Composition-conditioned Crystal GAN

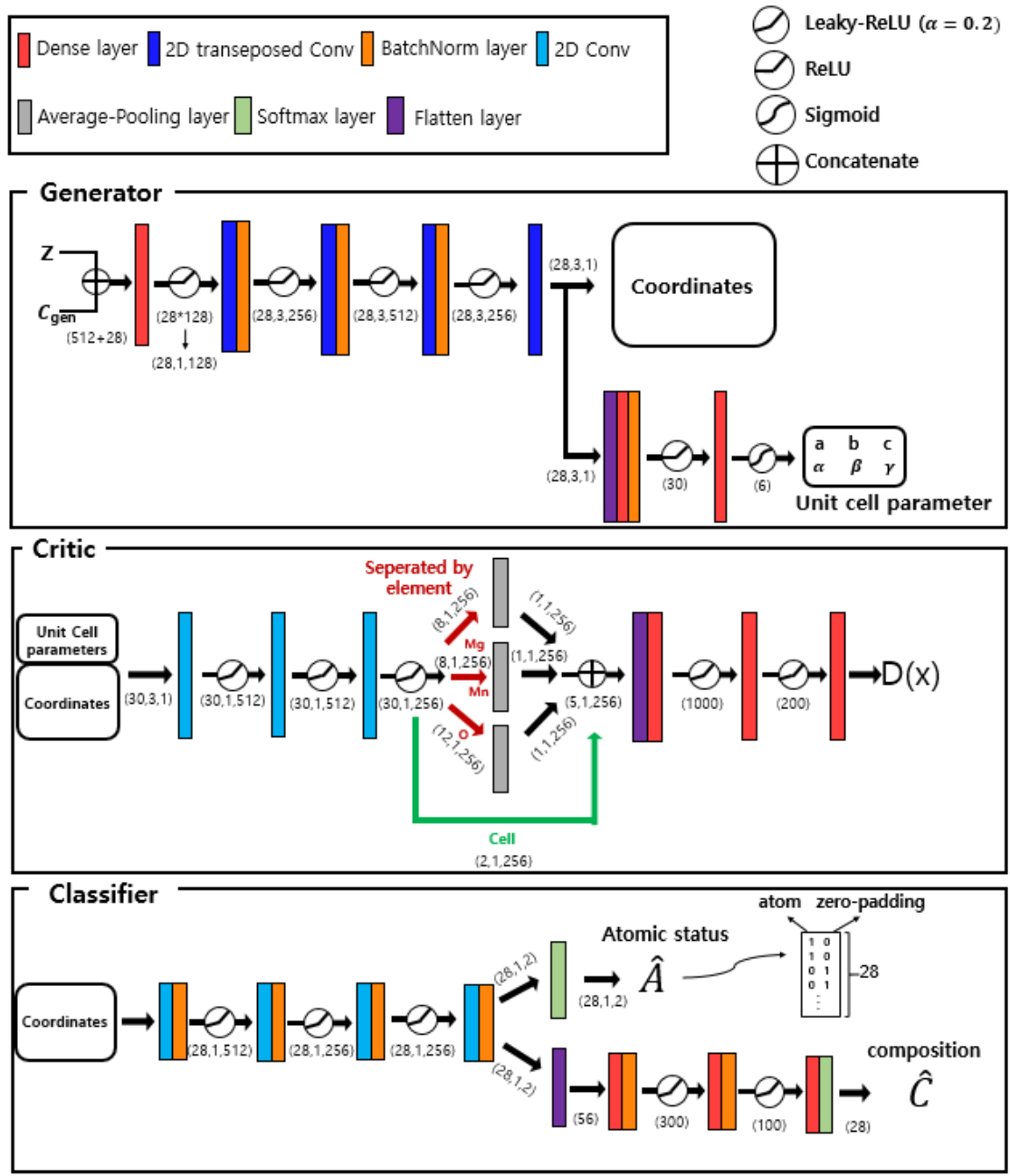

**Figure S6. Architecture of Composition-conditioned Crystal GAN**

**Table S1. Hyperparameters of Composition-conditioned Crystal GAN**

| Hyperparameter | | Value |
|---|---|---|
| Mini-batch size | | 32 |
| Adam optimizer | Learning rate | 0.0001 |
| | $\beta_1$ | 0.5 |
| | $\beta_2$ | 0.999 |
| Coefficient of Gradient penalty, $\lambda$ | | 10 |
| Coefficient of generated data in $L_{class-comp}$, $\lambda_1$ | | 1 |
| Coefficient of generated data in $L_{class-atom}$, $\lambda_2$ | | 1 |
| Coefficient of Composition Classification, $\lambda_c$ | | 0.3 |

**S2.2 Loss function of Composition-conditioned Crystal GAN**

In this work, we implemented variant of GAN called WGAN (Wasserstein GAN[67]) WGAN overcomes the shortcoming of GAN such as unstable training or mode collapse by using the Wasserstein distance between real and generated data distributions as the loss function. This loss function train the generator to create materials that are similar to the real materials. In addition, for improved training of WGAN, Gulrajani et al.[68] proposed additional term in loss function of WGAN, gradient penalty term. This regularizer term has enabled more stable training of WGAN. In detail, the loss function is

$$L_{WGAN} = \mathop{\mathbb{E}}_{\tilde{x} \sim \mathbb{P}_g}[D(\tilde{x})] - \mathop{\mathbb{E}}_{x \sim \mathbb{P}_r}[D(x)] + \lambda \mathop{\mathbb{E}}_{\hat{x} \sim \mathbb{P}_{\hat{x}}}[(\| \nabla_{\hat{x}} D(\hat{x}) \|_2 - 1)^2]$$

Where D is the critic function, $P_{\hat{x}}$ is sampling uniformly along straight lines between pairs of points sampled from the real data distribution, $P_r$ and the generator distribution, $P_g$, and λ is penalty coefficient, set to be 10. In order to train the generator to create materials with target property, the generator is trained together with the classifier with a loss function

$$L_{class-comp} = CE(C_{real}, \hat{C}_{real}) + \lambda_1 CE(C_{gen}, \hat{C}_{gen})$$

$$L_{class-atom} = CE(A_{real}, \hat{A}_{real}) + \lambda_2\, CE(A_{gen}, \hat{A}_{gen})$$

$$L_{class} = L_{class-atom} + \lambda_c L_{class-comp}$$

$$CE(t, x) = -\sum_{i}^{C} t_i \log(x_i)$$

where CE is cross entropy, $x_i$ is ith value of output value for classifier function, C is the number of classes and $t_i$ is ith target value. $C_{real}$, $\hat{C}_{real}$, $C_{gen}$, $\hat{C}_{gen}$ are true (no hat) and predicted (hat) property value of real (real subscript) and generated (gen subscript) materials. $\lambda_1$, $\lambda_2$, and $\lambda_c$ are generator coefficient in composition, generator coefficient in atomic state, and composition coefficient respectively.

## S3. Learning curve of the Composition-conditioned Crystal GAN

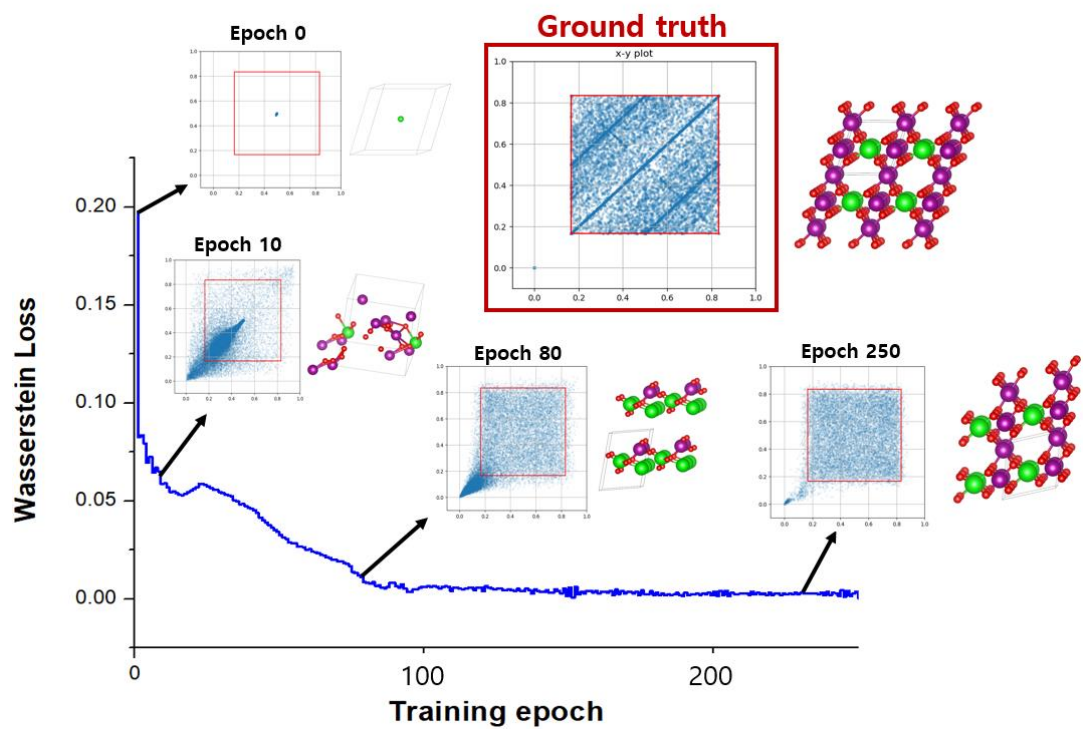

**Figure S7. The learning curve of Composition-conditioned Crystal GAN.** The Wasserstein loss as a function of the training epoch. The inset figures are scattered plot where scaled fractional coordinates of all atoms in generated structures are plotted. For clarity, only X-Y coordinates of atom positions are shown. The example of a generated structure according to the epoch is on the right side of the scatter plot. As the training progresses, generated atomic positions are becoming more and more similar to the ground truth representing the training dataset.

The learning process of our model is described in **Figure S7**, demonstrating that the Wasserstein distance converges to zero as training progresses and the generator can generate the data similar to real data. To help visualize it, we included the inset figures for scatter plot of all atoms' positions in randomly selected 1000 structures generated via Composition-conditioned Crystal GAN in the corresponding training epoch. In the ground-truth indicating training dataset, all atoms in the structures of training dataset are located in the red box except zero-paddings because the coordinates were rescaled. In the early stages of training, atoms were bound together and located in erroneous coordinates, however, as training progresses and Wasserstein distance converges to zero, the atomic and zero-padding coordinates become reasonable with the correct composition. The examples of converted to Mg-Mn-O structure were randomly selected from the generated structures in each corresponding training epoch and is located in the right side of each scatter plot.

## S4. Computational details

### S4.1 DFT calculations for the generated VO materials

For the comparison with iMatGen, we performed identical VASP[69] calculation method with Noh et al.[37] Also, phase diagram and energy above convex hull were calculated by the method described in methodology section of Noh et al[37] For all generated structures, we performed spin polarized GGA+U[70,71] calculations, with the same U parameter for V used in the Materials Projects database.[14] We relaxed both atomic positions and cell parameters using conjugate gradient descent method with convergence criteria of $1.0e^{-5}$ for energy and 0.05 eV/Å for force with 500 eV cut-off energy. To compare the phase stability among the generated structures for all generated materials we first used sparse reciprocal lattice grid with a grid spacing of 0.5 Å$^{-1}$. The formation energy (or formation enthalpy, $E_f$) was calculated using $E_{V_xO_y} - (xE_V - yE_O)/(x+y)$ (in eV/atom). Then, for a smaller set of materials that satisfy energy stability ($E_{hull} \leq 0.2$ eV/atom), we refined the formation energy calculations using a denser reciprocal lattice grid with grid spacing of 0.25 Å$^{-1}$.

### S4.2 DFT calculations for the generated MgMnO materials

For the comparison with high-throughput screening method, we performed identical VASP calculation method with Noh et al.[56] Also, phase diagram and energy above convex hull were calculated by the method described in methodology section of Noh et al.[56] For all generated structures, we performed spin-polarized PBE+$U$[70,71] calculations and PAW[72]–PBE pseudopotentials as implemented in the *ab initio* package, VASP, and we used 3.9 as U-value for Mn taken from Materials Project.[14] We relaxed both atomic positions and unit cell parameters using conjugate gradient descent method with convergence criteria of 1.0e-5 for energy and 0.05 for *eV*/Å force with 500 cut-off energy. To compare the phase stability among the generated structures for all generated materials, Brillouin zone is used with k-point densities at or larger than 500 k-points per atoms using the *Pymatgen*[64] package. Duplicates for the converged structures are removed using the *StructureMatcher* function implemented in *Pymatgen* package. After that, we performed the latter computations with dense k-space (i.e. Brillouin zone with k-point densities at or larger than 1000 k-points per atoms using the *Pymatgen* package).

### S4.3 Band gap calculations

We performed HSE[73] hybrid DFT functional implemented in VASP[69] with a mixing parameter of 0.2. For computational efficiency, a uniform reduction factor for the q-point grid of the exact exchange potential is applied (NKRED = 2) with gamma centered even number k-points (with a k-point densities at or larger than 1000 k-points per atoms).

## S5. Detailed statistics of the result

### S5.1 Materials generation statistics in Mg-Mn-O system

**Table S2. Numerical statistic of Mg-Mn-O polymorphs generated from Composition-conditioned Crystal GAN with compositions in MP.**

| Composition | E above hull≤200meV | E above hull≤80meV |
|---|---|---|
| MgMnO3 | 3 | - |
| MgMn2O4 | 35 | 8 |
| MgMn2O5 | 17 | - |
| MgMn3O6 | 15 | 2 |
| MgMn3O7 | 42 | - |
| MgMn4O6 | 52 | 7 |
| MgMn4O8 | 51 | 1 |
| Mg2Mn2O5 | 24 | 4 |
| Mg2Mn3O6 | 95 | 5 |
| Mg2Mn3O8 | 4 | - |
| Mg6MnO8 | 30 | 8 |

**Table S3. Numerical statistic of Mg-Mn-O polymorphs generated from Composition-conditioned Crystal GAN with compositions not in MP.**

| Composition | E above hull≤200meV | E above hull≤80meV |
|---|---|---|
| MgMn3O4 | 37 | 13 |
| MgMn3O5 | 39 | 7 |
| MgMn4O5 | 29 | 3 |
| MgMn4O7 | 26 | 3 |
| MgMn5O6 | 31 | 11 |
| Mg2MnO3 | 12 | 3 |
| Mg2MnO4 | 13 | 2 |
| Mg2Mn3O5 | 67 | 8 |
| Mg2Mn4O7 | 84 | 5 |
| Mg3MnO5 | 22 | 1 |
| Mg3Mn2O5 | 28 | 3 |
| Mg3Mn2O6 | 68 | 10 |
| Mg3Mn2O7 | 22 | - |
| Mg3Mn3O7 | 74 | 6 |
| Mg4MnO5 | 12 | 3 |
| Mg4MnO6 | 15 | 4 |
| Mg4Mn2O6 | 55 | 15 |
| Mg4Mn2O7 | 63 | 4 |
| Mg5MnO6 | 30 | 10 |
| Mg5MnO7 | 26 | 2 |

**S5.2 Structure comparison**

We first note that out of 28 Mg-Mn-O phases identified as promising photoanode materials, 14 phases correspond to new composition not included in database, meaning that the newly found structures are completely new and cannot be obtained from simple structural distortions. For the remaining 14 phases in 8 existing compositions, we estimated the similarity of the structures using the local structure order parameters with the 'Structure Matcher' function implemented in Pymatgen python package. (see **Figure S8**)

In **Figure S8**, we listed the 14 newly identified Mg-Mn-O structures with the most similar structures in our database based on the calculated dissimilarity values cited above. Indeed, we find that 5 of 14 identified structures (red boxed) are classified as the same structures to those in our database. For the other 9 structures, however, the generated structures seem to have very different structural motifs with large structural dissimilarity values compared those in the database. Thus, we believe the 23 out of 28 identified structures correspond to either completely new (with new compositions) or very different (with large dissimilarity values) structural motifs that are not in the databases.

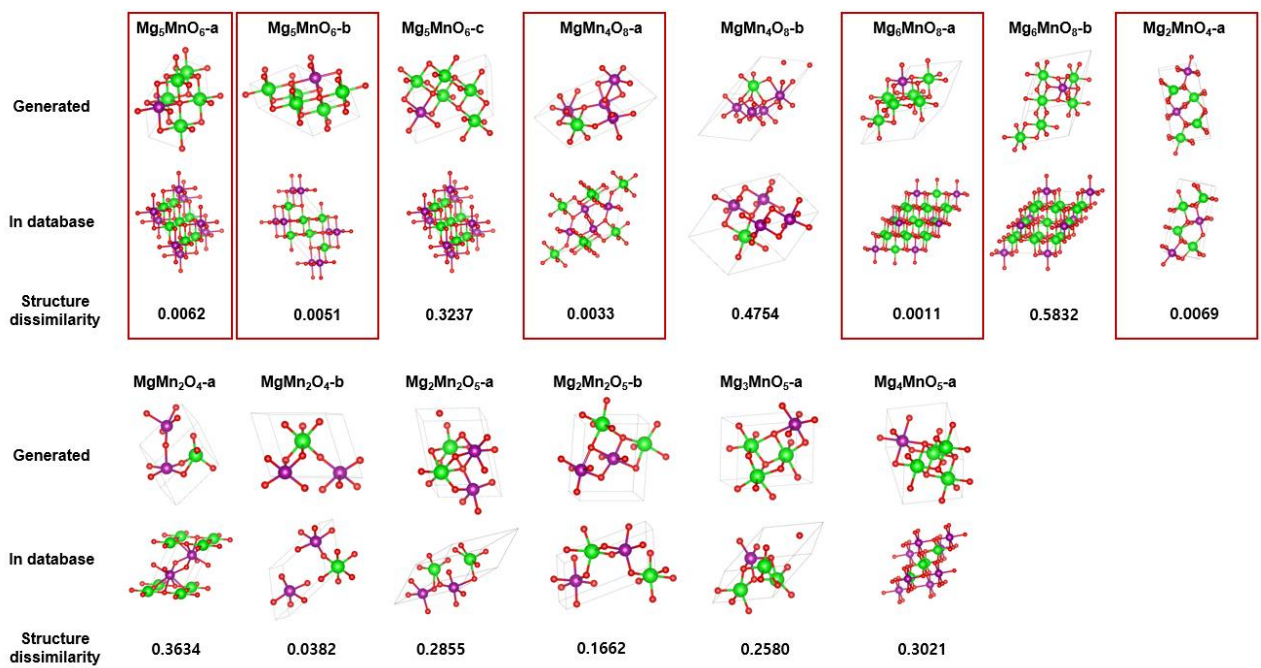

**Figure S8. The result of comparison of 14 generated structures considered as promising photoanode materials and database structures having same chemical composition.** In first row, 14 generated structures are located and in second row, the database structures most similar with corresponding generated structure are located. Structure dissimilarity in third row was value calculated between two structures in same column.

**S5.3 Examples of promising photo-anode materials**

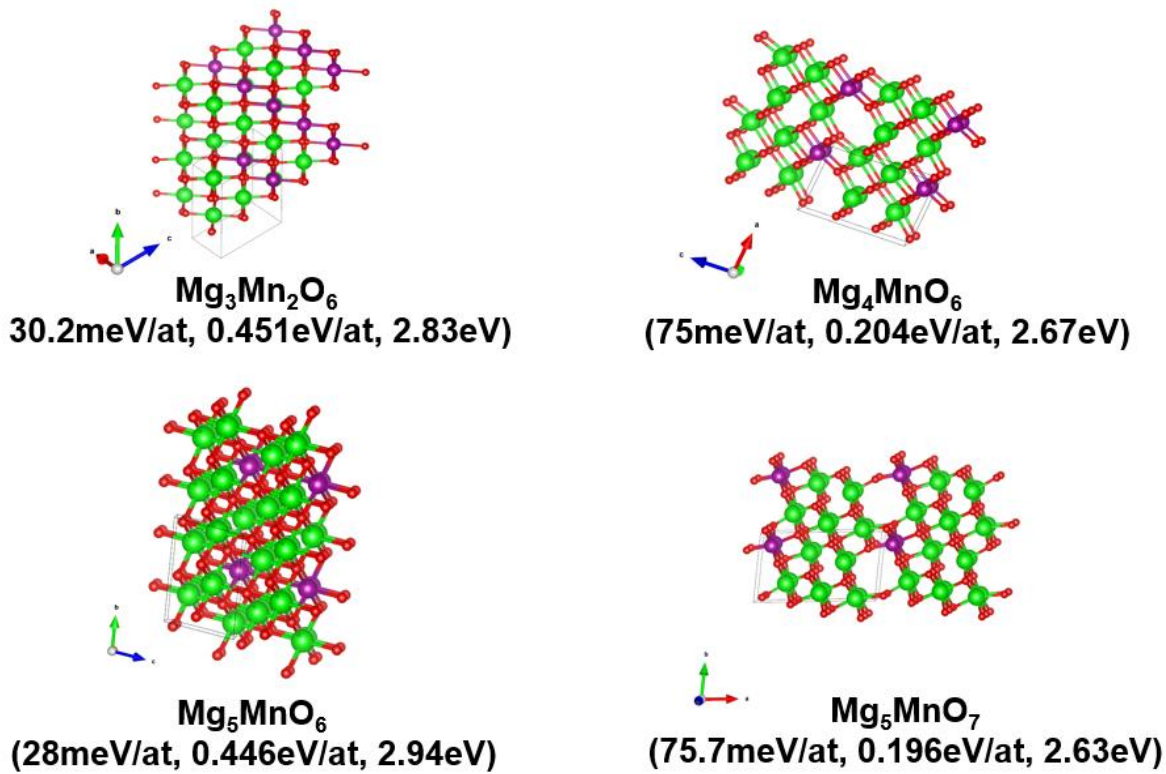

Mg$_3$Mn$_2$O$_6$
30.2meV/at, 0.451eV/at, 2.83eV)

Mg$_4$MnO$_6$
(75meV/at, 0.204eV/at, 2.67eV)

Mg$_5$MnO$_6$
(28meV/at, 0.446eV/at, 2.94eV)

Mg$_5$MnO$_7$
(75.7meV/at, 0.196eV/at, 2.63eV)

**Figure S9. Examples of promising photo-anode materials with composition not in MP.** For each structure, the energy above the convex hull, pourbaix stability, and bandgap energy are also shown below the structure.

## S6. V.O system

### S6.1 V.O dataset

In V-O system, $l_{Max}^{V}$ is 8 and $l_{Max}^{O}$ is 12. V-O initial dataset is constructed by using the elemental substitution for the binary compounds existing in the Material Project (MP) database, and removed duplicates. From this, we get a total of 1396 unique structures with 86 compositions for the V-O system. After data augmentation with the same way in Mg-Mn-O system, totaling 86,000 for V-O training dataset were constructed.

### S6.2 V.O results

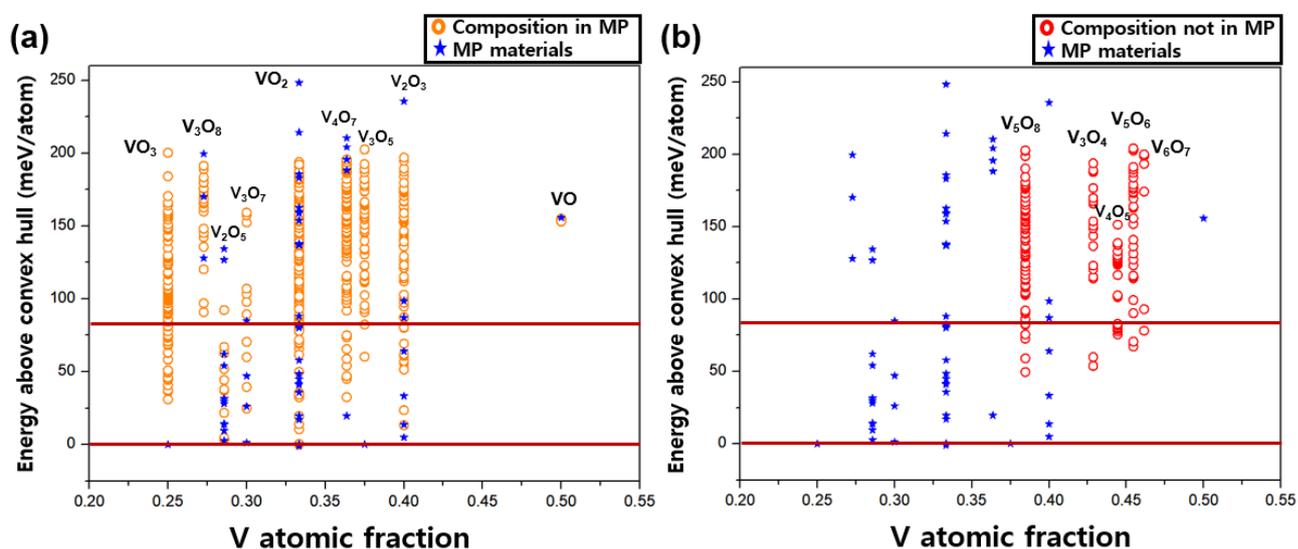

**Figure S10. DFT calculated formation energies for generated VO polymorphs.** (a) DFT calculated formation energies for the generated VO polymorphs with composition existing in MP (b) DFT calculated formation energies for the generated VO polymorphs with composition not in MP. Orange circles and Red circles are generated materials with composition existing in MP and not in MP respectively. Blue stars correspond to the materials of the V-O database in MP. The horizontal red lines represent 80meV/atom and 0meV/atom respectively.

To validate that our model can generate stable and realistic polymorphs of VxOy with given composition condition, we generated VxOy compounds of several known compositions in MP (see **Figure S10**), and screened them with their formation energy. We sampled structures according to these compositions and selected the structures only if the given composition condition ($C_{gen}$) is exactly reconstructed by the classifier ($\hat{C}_{gen}$). Also, we removed the structures if atoms are too close to each other. From this post-process, 300 structures in each composition thus totaling 5400 structures in 18 compositions were sampled, and further DFT calculations are performed to compute stability (i.e. the energy above the convex hull, $E_{hull}$) of the generated materials as shown in **Figure S10**. And for energy comparison with materials actually existing in MP, the formation energies of them were also plotted in **Figure S10**. Here, three main results are noteworthy:

(1) As shown in **Figure S10a**, 562 of unique and entirely new structures are predicted as theoretically meta-stable (i.e. $E_{hull} \leq 200$ meV/atom) among the total 4800 structures indicating that the proposed Composition-conditioned Crystal GAN can effectively generate stable and new materials. Furthermore, considering that 80 percent of the experimentally known identified sulfides and oxides were within this criterion[62] 91 unique structures are predicted as potentially synthesizable (i.e. $E_{hull} \leq 80$ meV/atom).

(2) Among 53 V-O polymorphs of MP which we compared with our generated results energetically in **Figure S10a** (blue stars), 15 materials have the same number of atoms in unit cell as the materials generated by our model. Although we randomly generated new materials using the trained model, 7 of 15 V-O polymorphs of MP with compositions which we selected to generate polymorphs are successfully re-discovered. Here, we expect the latter success ratio to increase further if we increase the number of samples.

(3) From comparison to the generated materials from iMatGen[37] (see **Figure S10b**), 40 percent of previous discovery was generated within the range of $E_{hull} \leq 200$ meV/atom by our framework (see **Figure S10b**). It is notable that for $V_3O_4$ our model generates more stable materials than the most stable previous discovery from iMatGen. In addition, in the case of $V_6O_7$, while iMatGen does not find stable materials in the range of $E_{hull} \leq 80$ meV/atom, our model can successfully generate new and stable polymorphs showing the effectiveness of our model to explore unseen chemical space.

**Table S4. Comparison with the results of iMatGen**[37] Each value in the table is the number of structures. Considering structures with $E_{hull}$ less than 200 meV/atom, 13 of 33 structures generated via iMatGen are composed in structures generated via Composition-conditioned Crystal GAN.

| Composition | Proposed Crystal GAN model | iMatGen | Identical structure |
|---|---|---|---|
| $V_3O_4$ | 20 | 10 | 3 |
| $V_4O_5$ | 3 | 3 | 1 |
| $V_5O_6$ | 32 | 8 | 3 |
| $V_5O_8$ | 75 | 10 | 4 |
| $V_6O_7$ | 6 | 2 | 2 |

**S7. The effect of data augmentation in generative model.**

To clarify that the data augmentation is actually affect the critic network to indirectly learn invariance under symmetric operations, we implemented additional experiment. To identify the effect of data augmentation and the minimum number of required data augmentation, we trained addition four independent models by changing the number of applied data augmentation as following:

-'No Aug': No augmentation (1,253 data points),

-'Aug1000': 1,000 structures per composition (112,000 data points total),

-'Aug2000': 2,000 structures per composition (224,000 data points total),

-'Aug3000': 3,000 structures per composition (336,000 data points total)

Next, we made a separate test dataset for 25 compositions. For each composition, there is 1 unique structure and 999 structures are augmented by applying the same procedure in the manuscript, yielding a total of 25,000 data points (25 x 1,000 data for each composition). To quantify the effect of data augmentation, we computed the difference of the output of the critic network between the base structure and the augmented structures. Here, the model would learn the invariance under the symmetric operations if the computed difference values are close to zero as shown in below figure. To be specific, in 'No Aug' case, there are relatively large deviation from the base structure (i.e. red line) compared to the other augmented models because the critic network cannot learn additional information from augmented structures. Notably, for the other 3 cases (i.e. 'Aug1000', 'Aug2000' and 'Aug3000'), the distribution is centered at zero (i.e. red line) indicating that the model can learn invariance under the symmetric operations from the augmented structures. Therefore, we can roughly estimate that augmenting 1000 data points per composition (in this work) would be sufficient.

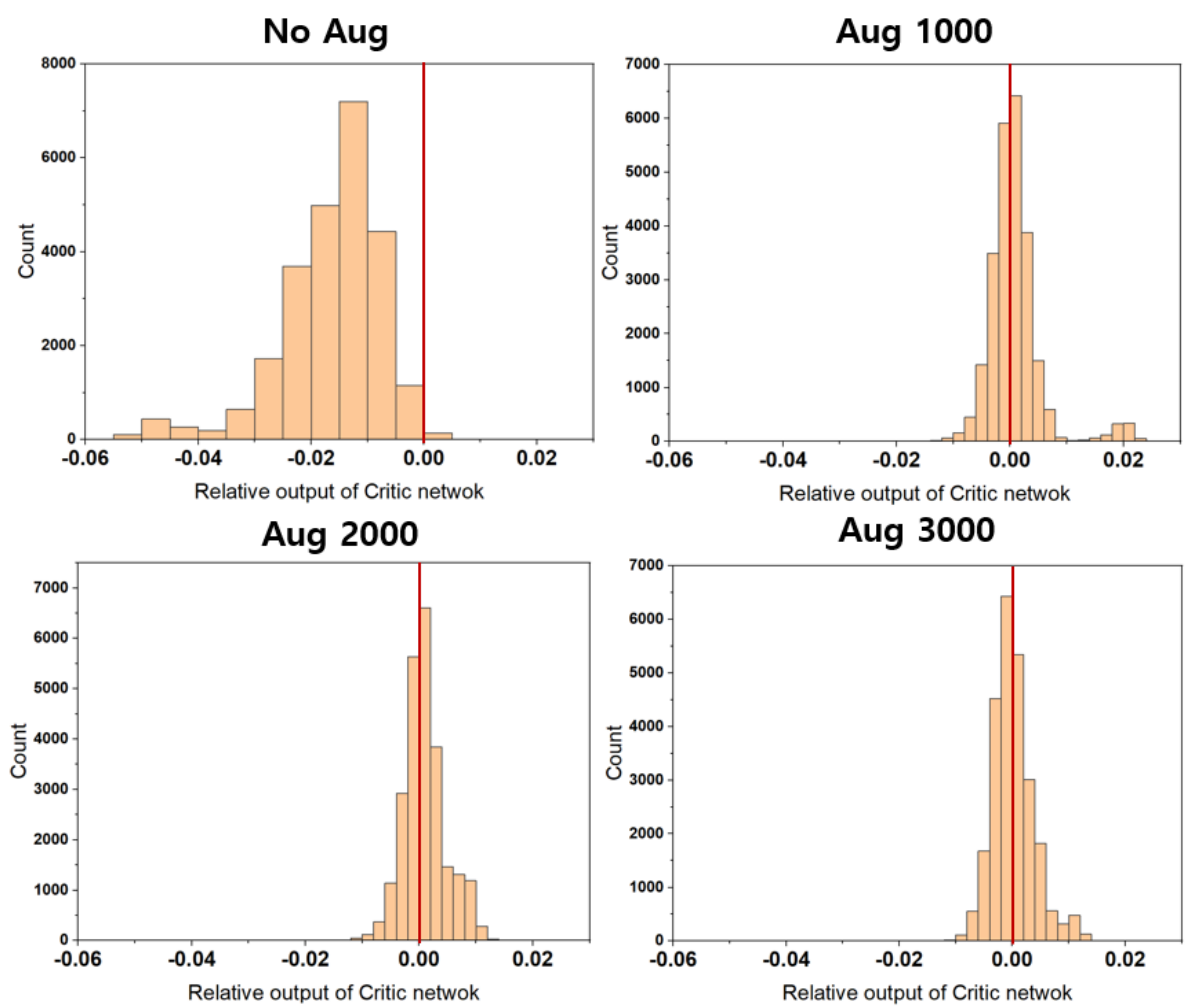

**Figure S11. Distribution of the difference of the output of critic network between the base structure and augmented structures.**